\documentclass[prl,twocolumn,superscriptaddress,showpacs]{revtex4}

\usepackage{graphicx}
\usepackage{amsmath}
\usepackage{amssymb}

\begin{document}

\newcommand{\ud}{{\mathrm d}}
\newcommand{\sech}{\mathrm{sech}}

\newcommand{\bs} {\boldsymbol}

\newcommand{\diff} {\mathrm{d}}
\newcommand{\f} {\frac}

\newcommand{\ga} {\alpha}
\newcommand{\gb} {\beta}
\newcommand{\gc} {\gamma}

\newcommand{\MB} {\mathrm{M}}
\newcommand{\J}  {\mathrm{J}}
\newcommand{\MJ}  {\mathrm{MJ}}

\newcommand{\Obs}  {\mathrm{O}}

\newcommand{\tot}  {\mathrm{tot}}
\newcommand{\Z}  {\mathcal{Z}}
\newcommand{\kB} {k_\mathrm{B}}


\title{Thermal equilibrium and statistical thermometers in special relativity}


\author{David Cubero}
\affiliation{F\'{\i}sica Te\'orica, Universidad de Sevilla, Apartado
  de Correos 1065, Sevilla 41080, Spain}
\author{Jes\'us Casado-Pascual}
\affiliation{F\'{\i}sica Te\'orica, Universidad de Sevilla, Apartado
  de Correos 1065, Sevilla 41080, Spain}
\author{J\"orn Dunkel}
\affiliation{Institut f\"ur Physik, Universit\"at Augsburg,
  Theoretische Physik I, Universit\"atsstra\ss{}e, D-86135 Augsburg, Germany}
\author{Peter Talkner}
\affiliation{Institut f\"ur Physik, Universit\"at Augsburg,
  Theoretische Physik I, Universit\"atsstra\ss{}e, D-86135 Augsburg, Germany}
\author{Peter H\"anggi}
\email{peter.hanggi@physik.uni-augsburg.de}
\homepage{www.physik.uni-augsburg.de/theo1/hanggi/}
\affiliation{Institut f\"ur Physik, Universit\"at Augsburg,
  Theoretische Physik I, Universit\"atsstra\ss{}e, D-86135 Augsburg, Germany}
\affiliation{Department of Physics, National University of Singapore,
Singapore 117542, Republic of Singapore}

\date{\today}

\begin{abstract}
There is an intense debate in the recent literature about the correct generalization of Maxwell's velocity distribution in special relativity. The most frequently discussed candidate distributions include the J\"uttner function as well as modifications thereof. Here, we report results from fully relativistic one-dimensional (1D) molecular dynamics (MD) simulations that resolve the ambiguity. The numerical evidence unequivocally favors the J\"uttner distribution. Moreover, our simulations illustrate that the concept of \lq thermal equilibrium\rq\space extends naturally to special relativity only if a many-particle system is spatially confined. They make  evident that \lq temperature\rq\space can be statistically defined and measured in an observer frame independent way. 
\end{abstract}


\maketitle

At the beginning of the last century it was commonly accepted that the one-particle velocity distribution of a dilute gas in equilibrium is described by the Maxwellian probability density function~(PDF)
\begin{equation}
\label{eq:maxwell}
f_\MB({\bs  v};m,\gb)=
\left(\frac{\gb m}{2\pi}\right)^{d/2} 
\exp\biggl(-\frac{\gb m \bs v^2}{2}\biggr)
\end{equation}
[$m$ is the rest mass of a gas particle, $\bs v\in \mathbb{R}^d$ the velocity, $T=(\kB \gb)^{-1}$ the temperature, $\kB$ the Boltzmann constant, $d$ the space dimension; throughout, we adopt natural units such that the speed of light $c=1$]. When Einstein~\cite{1905Ei_1,1905Ei_2} had formulated the theory of special relativity (SR) in 1905, Planck and others noted immediately that $f_\MB$ is in conflict with the fundamental relativistic postulate that velocities cannot exceed the light speed~$c$. A first solution to this problem was put forward by J\"uttner~\cite{1911Ju}. Starting from a maximum entropy principle, he proposed the following relativistic generalization of Maxwell's PDF:     
\begin{equation}
\label{eq:juttner}
f_\J({\bs  v};m,\gb_\J)
=\frac{m^{d}}{Z_\J}\gamma(\bs v)^{2+d}\exp[-\gb_\J m\gamma(\bs v)], \qquad 
|\bs v|<1
\end{equation}
[\mbox{$Z_\J=Z_\J(m,\gb_\J,d)$} is the normalization constant, $E=m\gc(\bs v)=(m^2+\bs p^2)^{1/2}$ the relativistic particle energy, $\bs p=m\bs v\gc(\bs v)$ the  momentum with Lorentz factor $\gamma(\bs v)=(1-\bs v^2)^{-1/2}$].  J\"uttner's distribution~\eqref{eq:juttner} became widely accepted among theorists during the first three quarters of the 20th century~\cite{1921Pa,1957Sy,1963Is,1971TeWe,1980DG} -- although a rigorous microscopic derivation is lacking due to the difficulty of formulating a relativistically consistent Hamilton mechanics of interacting particles~\cite{1949WhFe,1963CuJoSu,1965DaWi,1978Ko_b,1984MaMuSu}. Doubts about the J\"uttner function $f_\J$ began to arise in the 1980s, when Horwitz et al.~\cite{1981HoScPi,1989HoShSc} proposed a \lq manifestly covariant\rq\space relativistic Boltzmann equation, whose stationary solution differs from Eq.~\eqref{eq:juttner} and, in particular, predicts a different mean energy-temperature relation in the ultrarelativistic limit $T\to\infty$~\cite{2005Sc}. Since then, partially conflicting results and proposals from other authors~\cite{2002Ka,2006Le,2007DuHa,silva:057101,2007DuTaHa_2} have led to an increasing confusion as to which distribution actually represents the correct generalization of the Maxwellian~\eqref{eq:maxwell}. For  example, a recently discussed alternative to Eq.~\eqref{eq:juttner} is the \lq modified\rq\space J\"uttner function~\cite{2006Le,2007DuHa}
\begin{equation}
f_\MJ({\bs  v};m,\gb_\MJ)=
\frac{m^{d}}{Z_\MJ}\frac{\gamma(\bs v)^{2+d}}{m\gamma(\bs v)}
\exp[-\gb_\MJ m\gamma(\bs v)].
\label{eq:modjuttner}
\end{equation}
The distribution~\eqref{eq:modjuttner} can be obtained e.g. by combining a maximum relative entropy principle and Lorentz symmetry~\cite{2007DuTaHa_2}. Compared with $f_\J$ at the same parameter values \mbox{$\gb_{\J}=\gb_\MJ\lesssim 1/m$}, the modified PDF $f_\MJ$ exhibits a significantly lower particle population in the high energy tail because of the additional $1/E$-prefactor. 
\par
Identifying the correct relativistic equilibrium velocity distribution is essential for the proper interpretation of present and future experiments in high energy and astrophysics~\cite{1998ItKoNo,2006DiDrSh,2006RaGrHe,2006HeGrRa}. Examples include the application of relativistic Langevin equations~\cite{1997DeMaRi,2005Zy,2005DuHa} to heavy ion collision experiments~\cite{2006HeGrRa,2006RaGrHe}, thermalization processes in ultra-relativistic plasma beams~\cite{2006DiDrSh}, or the relativistic Sunyaev-Zel'dovich (SZ) effect~\cite{1998ItKoNo}, describing the distortion of the cosmic microwave background (CMB) radiation spectrum due to the interaction of CMB photons with  hot electrons in clusters of galaxies~\cite{1972SuZe,1984SZNature,1993SZNature}. The predicted strength of these spectral distortions and the cosmological parameters inferred from the SZ effect depend sensitively on the assumed electron velocity distribution~\cite{1998ItKoNo}.
 
\paragraph*{Relativistic MD simulations.--} 
To resolve the uncertainty about the relativistic equilibrium velocity PDF, we performed fully relativistic one-dimensional (1D) molecular dynamics (MD) simulations. The restriction to the 1D case is inevitable if one wants to treat localized particle interactions in a relativistically consistent manner, cf. remarks below. In our computer experiments we simulated the dynamics of classical, impenetrable point-particles with elastic point-like binary collisions, employing an algorithm similar to those of Alder and Wainwright~\cite{1959AlWa} and Masoliver and Marro~\cite{1983MaMa}. The basic time step of the algorithm involves three partial tasks: $(i)$ determine the next collision event $(x_c,t_c)$; $(ii)$ evolve the system up to time $t_c$; $(iii)$ calculate the momenta after the collision. The third task is solved as follows: If two particles $A$ and $B$ meet at the space-time point $(x_c,t_c)$, then they exchange momentum according to the relativistic energy momentum conservation laws, 
\begin{eqnarray}\label{eq:conservation}
\begin{split}
p_A+p_B&=\hat{p}_A+\hat{p}_B\\
E(m_A,p_A)+E(m_B,p_B)&= E(m_A,\hat p_A)+E(m_B,\hat p_B).
\end{split}
\end{eqnarray}
Here $p=mv\gc(v)$ is the relativistic momentum, and $E(m,p)=(m^2+p^2)^{1/2}$ the energy; hat-symbols denote quantities after the  collision. Given the momenta $(p_A, p_B)$ before the collision, the conservation laws~\eqref{eq:conservation} determine the momenta $(\hat p_A, \hat p_B)$ after the collision by~\cite{2007DuHa}
\begin{eqnarray}\label{eq:col}
 \begin{split}
 \hat{p}_A=\gc(v_0)^2 [2v_0 E(m_A,p_A)-(1+v_0^2)p_A], \\
 \hat{p}_B=\gc(v_0)^2 [2v_0 E(m_B,p_B)-(1+v_0^2)p_B],
 \end{split}
 \end{eqnarray}
where $v_0=(p_A+p_B)/[E(m_A,p_A)+E(m_B,p_B)]$ is the collision-invariant, relativistic center-of-mass velocity of the two particles. By assuming strictly localized, point-like pair interactions, one may avoid the introduction of fields which are required when considering relativistic particle interactions-at-a-distance (the interested reader may wish to consult the original papers of Wheeler and Feynman~\cite{1949WhFe}, Currie et al.~\cite{1963CuJoSu}, and Van Dam and Wigner~\cite{1965DaWi,1966DaWi}, who discuss in detail the difficulties associated with classical particle-particle interactions in SR). However, considering point-like localized interactions is expedient in the 1D case only; in higher space dimension the collision probability would become zero, thus preventing the system from equilibration. Moreover, if two colliding particles carry the same rest masses then elastic 1D collisions merely interchange their velocities; hence, elastic binary collisions are not able to drive a 1D \emph{one-component} gas to equilibrium.  In our simulations we considered a \emph{two-component} mixture, consisting of $N_1$ light particles having equal masses $m_1$, and $N_2$ particles with equal masses $m_2>m_1$. The motion of the $N=N_1 + N_2$ particles was restricted to the 1D interval $[0,L]$, assumed to be stationary in the lab frame $\Sigma$. The results presented below refer to elastic reflections at the boundaries; however, we found that periodic boundary conditions yield identical outcomes if the total initial momentum was chosen to be zero in $\Sigma$. Generally, our simulations mimic a relativistic microcanonical ensemble, since the total initial energy $E_\tot$ in $\Sigma$ is conserved in the microscopic collision processes. The above conventions define the simplest interacting model system that $(i)$ complies with all principles of SR, $(ii)$ does not require the introduction of interaction fields,  $(iii)$ can be simulated without further approximation, and $(iv)$ exhibits a universal stationary equilibrium state. Hence, this model systems provides an optimal test case for probing the predictions of different relativistic kinetic theories by means of numerical experiments~\cite{1957Sy,1963Is,1971TeWe,1980DG,1981HoScPi,2005Sc}. Moreover, as we shall see below, it helps to clarify longstanding controversial questions regarding the definition and meaning of \lq temperature\rq\space (i.e. thermometers) and \lq thermal equilibrium\rq\space in SR.

\begin{figure}[h]
\includegraphics[width=7.5cm,angle=0]{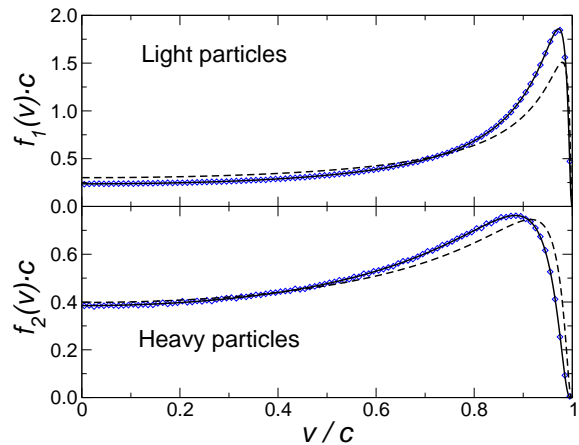}
\caption{
\label{fig:1} 
Equilibrium PDFs in the lab frame $\Sigma$: Numerically obtained one-particle velocity PDFs ($\diamond$) based on a simulation with $N_1=5000$ light particles of mass $m_1$ and $N_2=5000$ heavy particles with mass $m_2=2m_1$. The mean energy per particle in $\Sigma$ is $\epsilon=E_\tot/(N_1+N_2)=2.5 m_1c^2$.  The solid curves correspond to J\"uttner functions~\eqref{eq:juttner} with same parameter $\gb_\J=0.702\, (m_1c^2)^{-1}$, but different particle masses, respectively.  Dashed lines show the corresponding modified distribution~\eqref{eq:modjuttner} with $\gb_\MJ=0.402\,(m_1c^2)^{-1}$. As the distributions are symmetric with respect to the origin, only the positive velocity axis is shown. The simulation data is consistent with the standard J\"uttner distribution \eqref{eq:juttner}, and thus provides evidence against the modified distribution~\eqref{eq:modjuttner}.
}
\end{figure}

\paragraph*{Numerical results.--} 
In order to identify the stationary one-particle velocity distributions 
for the light and heavy particles, respectively, we waited until the 1D two-component gas had approached the equilibrium state (typically, after $10^2$ collisions per particle). Then the particle velocities were measured $\Sigma$-simultaneously, i.e., at equal times with respect to the resting lab frame $\Sigma$. To increase the sample size we repeated this procedure several times during a simulation run and collected the data into a single histogram.    
An example is shown in Fig.~\ref{fig:1}, based on a simulation with $N=10000$ particles ($N_1=N_2=5000$, $m_2=2m_1$). Each particle had been given a random initial position $x_i(0)\in [0,L]$ and a random initial velocity $v_i(0)=\pm 0.8$,  corresponding to a mean energy per particle $\epsilon=2.5 m_1$. As evident from Fig.~\ref{fig:1}, for both particle species the numerically obtained one-particle PDFs ($\diamond$) are in very good agreement with the standard J\"uttner function $f_\J$ (solid line), and differ significantly from the modified distribution~$f_\MJ$ (dashed lines). The same result was found for $N_1\ne N_2$.
\par
The distribution parameters $\gb_{\J/\MJ}$ were determined from the initial energy by means of the following consideration:
If the particle numbers $N_1$ and $N_2$ are sufficiently large (thermodynamic limit), then the one-particle PDFs in the lab frame $\Sigma$ are expected to converge to either $f_\J$ from Eq.~\eqref{eq:juttner} or $f_\MJ$ from Eq.~\eqref{eq:modjuttner}. Generally, the mean relativistic energy value $\mu$ of a one-particle PDF $f(\bs v;m,\gb)$ is given by
 \begin{equation}
 \mu(m,\gb)=\int_{\{|\bs v|<1\}}\diff^d\bs v\; 
 f(\bs v;m,\gb)\;m\gc(\bs v).
 \end{equation}
 Assuming $(i)$ that an equilibrium state exists where both species can be described by the same value  $\gb$, and $(ii)$ that for a gas in equilibrium the mean energy per particle is the same for particles \emph{of the same species}, the total energy can be expressed as 
 \begin{equation}\label{eq:beta-constraint}
 E_\tot=N_1\; \mu(m_1,\gb) +N_2\; \mu(m_2,\gb). 
 \end{equation}
 In our case, the energy mean values of the two 1D candidate PDFs $f_\J$ and $f_\MJ$ read explicitly
 \begin{eqnarray}\label{eq:beta-constraint-2}
 \begin{split}
 \mu_\J(m,\gb_\J)&=m\f{K_0(\gb_\J m)+K_2(\gb_\J m)}{2K_1(\gb_\J m)},\\
 \mu_\MJ(m,\gb_\MJ)&=m \f{K_1(\gb_\MJ m)}{K_0(\gb_\MJ m)},
 \end{split}
 \end{eqnarray}
 with $K_n$ denoting the modified Bessel function of the second kind~\cite{AbSt72}. For each simulation run the parameter tuple $(E_\tot,N_1,N_2,m_1,m_2)$ is known. Hence, upon inserting them into Eqs.~\eqref{eq:beta-constraint} and \eqref{eq:beta-constraint-2}, these parameters uniquely determine the parameter value $\gb_{\J/\MJ}$ that is consistent with the chosen velocity PDF~$f_{\J/\MJ}$. 

\paragraph{Temperature and equilibrium.--} 
Most remarkably, in spite of the different particle masses the two numerically obtained velocity PDFs in Fig.~\ref{fig:1} are very well matched by J\"uttner functions~\eqref{eq:juttner} with the \emph{same} parameter $\gb_\J$. According to our simulations, this holds true with high accuracy for a wide range of initial conditions and mass ratios. Hence, the J\"uttner function does not only provide the best \lq fit\rq\space to the numerical data, it also yields a well-defined concept of \lq temperature\rq\space in~SR: Intuitively, the temperature $T$ is thought to be an intensive quantity that equilibrates to a common value if two or more systems are brought into contact with each other (i.e., may exchange different forms of energy). In our case, it is natural to consider the particle species as two different subsystems that may exchange energy via elastic collision processes. After a certain relaxation time, the combined system approaches a \lq thermodynamic equilibrium state\rq, where each subsystem is described by the same asymptotic, two-parametric velocity PDF $f_\J(v;m_i,\gb_\J)$, differing only via the rest masses~$m_i$. The commonly shared distribution parameter $\gb_\J$ may thus be used to \emph{define} a relativistic equilibrium temperature $T:=(\kB\gb_\J)^{-1}$. However, for this concept to be meaningful, a restriction of the accessible spatial volume is required -- be it by means of periodic boundary conditions, or by imposing reflecting walls. Otherwise, it cannot be expected that a many-particle system approaches a universal stationary state which is independent of the specific initial conditions. This observation has an important implication: Any (relativistic or non-relativistic) Boltzmann-type equation~\cite{1963Is,1980DG,1981HoScPi,1989HoShSc,Cercignani,2005LivRev,2005Sc} that gives rise to a universal stationary velocity PDF implicitly assumes the presence of a spatial confinement, thus singling out a preferred frame of reference.

\begin{figure}[t]
\includegraphics[width=7.5cm,angle=0]{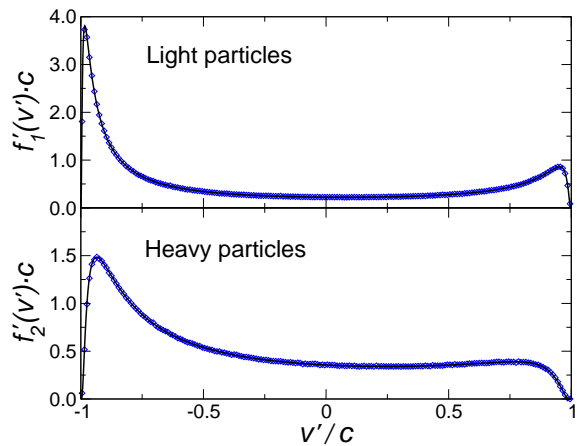}
\caption{
\label{fig:2} 
Equilibrium PDFs in a moving frame $\Sigma'$: Velocity PDFs as measured by an observer who moves with velocity $u=0.25c$ relative to the lab frame $\Sigma$. Parameter values are the same as in Fig.~\ref{fig:1}. The solid lines correspond to J\"uttner functions $f'_\J$ from Eq.~\eqref{eq:moving} with the same parameter $\gb_\J=0.702\, (m_1c^2)^{-1}$ as in Fig.~\ref{fig:1} and different masses $m_1$ and $m_2$, respectively.
}
\end{figure}

\paragraph{Moving observers and statistical thermometers.--} 
From our simulations we may further determine the equilibrium velocity distributions as seen from another frame $\Sigma'$ moving with velocity $u$ relative to the lab frame~$\Sigma$. Figure~\ref{fig:2} depicts the results for  $u=0.25$ and same simulation parameters as in Fig.~\ref{fig:1}. In contrast to Fig.~\ref{fig:1},  the numerical data points in Fig.~\ref{fig:2} were obtained by measuring velocities $\Sigma'$-simultaneously. The solid curves in Fig.~\ref{fig:2} correspond to the PDF
\begin{equation}\label{eq:moving}
f_\J'(v';m,\gb_\J,u)=
\frac{m\,\gamma(v')^{3}}{ Z_\J\,\gc(u)}
\exp[-\gb_\J\gc(u) m\gamma(v')\; (1+u v')]
\end{equation}
[$v'$ is the particle velocity in the moving frame~$\Sigma'$].  The PDF~\eqref{eq:moving} reduces to the J\"uttner function~\eqref{eq:juttner} for $u=0$; $f_\J'$~is obtained by using the fact that the one-particle phase space PDF, reading 
$$
\Phi_\J(x,p)=(Z_\J L)^{-1}\exp[-\gb_\J E(m,p)]\,\Theta(x)\,\Theta(L-x)
$$ 
in $\Sigma$, is a Lorentz scalar~\cite{1924Di,1969VK}; $\Theta$ is the Heaviside unit-step function. Due to the excellent agreement between the numerical simulations and Eq.~\eqref{eq:moving}, we may state more precisely: Two relativistic gas components are in \lq thermodynamic equilibrium\rq\space for any observer if their one-particle velocity PDFs are given by generalized J\"uttner functions~\eqref{eq:moving} with same parameters $\gb_\J$ \emph{and} $u$. Only in this case the net energy transfer between the different gas components in the container vanishes. 
Last but not least, the above results shed light on a longstanding, highly debated question~\cite{1966Fr,1967Wi,1967La,1967No,1967Wi_2,1996LaMa} originally posed by P. T. Landsberg~\cite{1966La}: 

\paragraph{Does a moving body appear cool? --} 
Evidently, the answer depends on the thermometers employed by different observers. Adopting, for the reasons discussed above, $T:=(\kB \gb_\J)^{-1}$ as a reasonable temperature definition, a moving observer with rest frame $\Sigma'$ can measure $T$ by exploiting the Lorentz invariant equipartition theorem~\cite{1967La}
\begin{eqnarray}\label{eq:thermometer}
\kB T=m\gc(u)^3\left\langle {\gc(v')\;(v'+ u)^2} \right\rangle',
\end{eqnarray}
where $u=-\langle v'\rangle'$, and averages $\langle\,\cdot\,\rangle'$ are taken  $\Sigma'$-simultaneously. We verified the validity of Eq.~\eqref{eq:thermometer} explicitly by using simulation data obtained for different values of $u$. Hence, Eq.~\eqref{eq:thermometer} defines a Lorentz invariant gas thermometer on a purely microscopic basis. Put differently, this intrinsic statistical thermometer determines the proper temperature of the gas by making use of simultaneously measured particle velocities only; thus, \emph{moving bodies appear neither hotter nor colder}.  Analogous considerations apply to the 2D/3D case.

\paragraph{Summary.--}
Fully relativistic MD simulations favor the J\"uttner distribution~\eqref{eq:juttner} as the correct relativistic one-particle equilibrium velocity distribution. The results are conclusive for the 1D case, and provide evidence against theories~\cite{1981HoScPi,1989HoShSc,2002Ka,2005Sc,2006Le,2007DuHa} that predict other distributions.  Further, our simulations corroborate Landsberg's hypothesis~\cite{1966La,1967La} that the temperature of classical gaseous systems can be defined and measured in a Lorentz invariant way. The extension of the MD approach to higher space dimensions is nontrivial, due to the fundamental difficulty of treating 2D and 3D two-body collisions in a relativistically consistent manner~\cite{1949WhFe,1963CuJoSu,1965DaWi,1978Ko_b,1984MaMuSu}. As a first step, it should be carefully analyzed if and how specific semi-relativistic interaction models affect the 2D/3D equilibrium velocity distribution. 

\paragraph{Acknowledgements.--}
This research was supported by the Juan de la Cierva programe of the Ministerio de Ciencia y Tecnolog\'{\i}a (D.C.), the Direcci\'on General de Ense\~nanza Superior of Spain by Projects Nos. BFM2002-03822 (J.C.-P.), the Junta de Andalucia (J.C.-P. and D.C.). Financial support of the German Excellence Initiative via the \lq\lq Nanosystems Initiative Munich (NIM)\rq\rq\space is gratefully acknowledged (P. H.).

\bibliography{References}


\end{document}